\documentclass[superscriptaddress,twocolumn]{revtex4}
\usepackage{graphicx}

\begin{document}

\title{Separation of VUV/UV photons and reactive particles in the effluent of a He/O$_2$ atmospheric pressure plasma jet}

\author{S. Schneider}
\address{Coupled plasma-solid state systems, Fakult\"at f\"ur
  Physik und Astronomie, Ruhr-Universit\"at Bochum,
  Universit\"atsstr. 150, 44780 Bochum, Germany}

\author{J.-W. Lackmann, F. Narberhaus, J. E. Bandow}
\address{Mikrobiologie, Fakult\"at f\"ur
  Biologie, Ruhr-Universit\"at Bochum,
  Universit\"atsstr. 150, 44780 Bochum, Germany}

\author{B. Denis}
\address{Institute for Electrical Engineering and Plasma Technology,
Ruhr-Universit\"at Bochum,
  Universit\"atsstr. 150, 44780 Bochum, Germany}

\author{J. Benedikt}
\address{Coupled plasma-solid state systems, Fakult\"at f\"ur
  Physik und Astronomie, Ruhr-Universit\"at Bochum,
  Universit\"atsstr. 150, 44780 Bochum, Germany}

\begin{abstract}

Cold atmospheric pressure plasmas can be used for treatment of living tissues or for inactivation
of bacteria or biological macromolecules. The treatment is usually characterized by a combined
effect of UV and VUV radiation, reactive species, and ions. This combination is usually beneficial
for the effectiveness of the treatment but it makes the study of fundamental interaction mechanisms
very difficult. Here we report on an effective separation of VUV/UV photons and heavy reactive
species in the effluent of a micro scale atmospheric pressure plasma jet ($\mu$-APPJ). The
separation is realized by an additional flow of helium gas under well-defined flow conditions,
which deflects heavy particles in the effluent without affecting the VUV and UV photons. Both
components of the effluent, the photons and the reactive species, can be used separately or in
combination for sample treatment. The results of treatment of a model plasma polymer film and
vegetative \emph{Bacillus subtilis} and \emph{Escherichia coli} cells are shown and discussed. A
simple model of the He gas flow and reaction kinetics of oxygen atoms in the gas phase and at the
surface is used to provide a better understanding of the processes in the plasma effluent. The new
jet modification, called X-Jet for its appearance, will simplify the investigation of interaction
mechanisms of atmospheric pressure plasmas with biological samples.

\end{abstract}

\maketitle

\section{Introduction}

Cold atmospheric pressure plasmas (CAP) are increasingly in the focus of researchers investigating
their possible applications in medicine or the food packaging industry
\cite{Kong2009,Stoffels2008}. CAP jets are able to inactivate bacteria \cite{Daeschlein2010}, fungi
\cite{Daeschlein2010a}, or bio-macromolecules \cite{Deng2007} and offer an alternative to standard
sterilization methods, where thermo-labile and vacuum-sensitive objects (plastics or living
tissues) have to be treated. In addition, several studies indicate that CAP treatment can
accelerate wound healing or influence cancer cells \cite{Fridman2007,Vandamme2010}. These plasmas
can be filamentary or glow dielectric barrier discharges (DBD) with few kV voltages, usually pulsed
at kHz frequencies \cite{Dobrynin2009,Morfill2009}, DBD based jets with so called plasma bullets
\cite{Cao2009,Laroussi2005,Robert2009}, or homogeneous glow discharges generated in RF-MW
frequencies in form of plasma jets \cite{Schulz-vonderGathen2007,Stoffels2008,Ehlbeck2011}. The
term "cold" refers to temperatures in the plasma effluent close to room temperature allowing direct
treatment for instance of skin or thermolabile polymers. These plasmas are operated in air or
air-containing gas mixtures. Alternatively they use only He with or without admixture of some
molecular gas (usually O$_2$). These plasmas produce positive and negative ions, (V)UV radiation,
and reactive radical species, which interact with the treated surface. The effects of different
plasma-generated species on the treated systems are a topic of current scientific discussions
\cite{Kong2009}. The role of reactive oxygen species (ROS) has been stressed by several authors as
a key factor influencing vegetative prokaryotic and eukaryotic cells \cite{Goree2006,Haehnel2010}.
More recently, the combination of ions and ROS has also been discussed
\cite{Dobrynin2009,Stoffels2008}.

Atmospheric pressure plasma jet (APPJ) sources operated with He with some addition of O$_2$ ($\leq$
1\%) are known to be efficient sources of ROS, particularly oxygen atoms, ozone molecules (O$_3$),
or singlet delta oxygen metastables O$_2$ (a$^1\Delta_g$). Measurements and modelling have been
reported for a coaxial jet with 1 cm diameter inner electrode and 1\,mm electrode gap
\cite{Jeong98}, parallel plate jet with 1\,mm electrode separation and 1\,mm electrode width
\cite{Schulz-vonderGathen2007}, or for sources with electrode width larger than 1\,mm
\cite{Schulz-vonderGathen2007,Laimer2006}. The plasma dynamics and plasma chemistry in these
discharges have also been modelled by several authors \cite{Waskoenig2010,Liu2010}. These works
show that densities of above mentioned ROS are around 10$^{15}$\,cm$^{-3}$ in the effluent of these
jets and can be tuned by adjusting O$_2$ concentration, applied power, gas flow, and jet-substrate
distance. This kind of plasma is a promising tool for treatment of living tissues or for
antibacterial treatment of surfaces at atmospheric pressure. The knowledge of ROS densities and,
therefore, also the fluxes could be used to evaluate quantitatively the effects of ROS for example
on vegetative bacteria. An unknown factor in these studies is the amount of VUV and UV photons,
which are produced next to ROS in the plasma. These photons propagate unabsorbed through He
atmosphere, irradiate the treated surface and can induce uncontrolled radiation damage. We report
in this article a method, in which plasma generated VUV and UV photons and heavy reactive particles
are effectively separated from each other and can be used separately for the surface treatment.

The manuscript is organized as follows. First, the microplasma jet used for the plasma generation
is described and its modification allowing the separation is introduced. Second, the fluid dynamic
model for the simulation of the gas flow and reaction kinetics of O atoms and ozone is described
and discussed. Third, the performance of the X-Jet is tested by measuring the light emission in the
115-875\,nm wavelength region and by analyzing the etching profiles of a model polymer film.
Finally, the study of the interaction of plasma effluent with bacteria is presented.

\section{Experimental Setup}

\begin{figure}
\centering
\includegraphics[width=8cm]{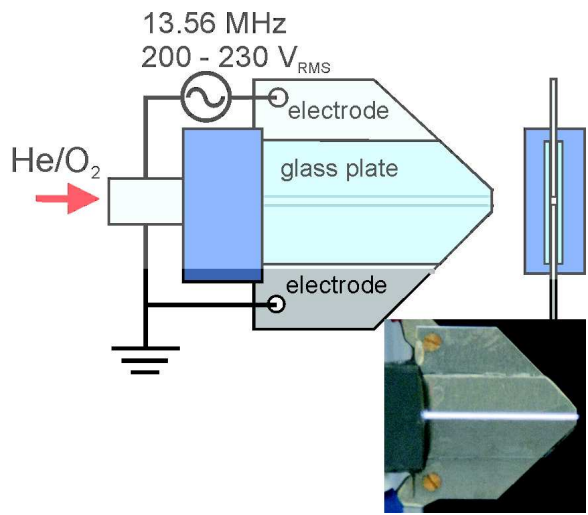}
\caption{Schematics of a $\mu$-APPJ source and its photograph during the operation with He/O$_2$
gas mixture.}\label{fig:APPJ}
\end{figure}

Two different jets are used in this work. The first plasma source is a parallel plate microscale
atmospheric pressure plasma jet ($\mu$-APPJ) and the second source is its modification with two gas
channels at the jet nozzle. The $\mu$-APPJ is a capacitively coupled microplasma jet consisting of
two stainless steel electrodes (length 30\,mm, thickness 1\,mm) with a separation of 1\,mm, and two
glass plates, which confine the inter electrode volume on the sides (cf. Fig.\ref{fig:APPJ}). One
electrode is connected to a power supply (13.56\,MHz,  applied root-mean-square voltage 200-230\,V,
absorbed power $<$ 1\,W) through a matching network and the other one is grounded. The volume of
the plasma is 1\,x\,1\,x\,30\,mm$^3$. The $\mu$-APPJ used here equals the $\mu$-APPJ described
elsewhere \cite{Knake08, Ellerweg2010}. The He gas flow through the jet is 1.4\,slm with a small
admixture of molecular oxygen (8.4\,sccm, 0.6$\%$). It has been shown that this microplasma jet is
a typical $\alpha$-mode discharge \cite{Schulz-vonderGathen2007} and quantitative measurement of O
and O$_3$ densities as function of O$_2$ concentration, applied power, and distance to the jet are
available \cite{Knake08, Ellerweg2010}. For example, the densities (concentrations) of O and O$_3$
measured by molecular beam mass spectrometry at 4\,mm distance from the jet under conditions used
in this work are 7$\times$10$^{14}$\,cm$^{-3}$ ($\sim$28\,ppm) and 5$\times$10$^{14}$\,cm$^{-3}$
($\sim$20\,ppm) respectively \cite{Ellerweg2010}. The concentration of ozone increases with the
distance and reaches $\sim$56\,ppm at 50\,mm. The substrate to be treated is placed perpendicular
to the jet axis at a distance of 4\,mm.

The gas temperature can be an issue for treatment of bacteria. The gas temperature in the effluent
has been measured for 1$\%$ of O$_2$ admixture and it was below 34$^{\circ}$C under these
conditions \cite{Schulz-vonderGathen2007}. The temperature is slightly higher at lower O$_2$
concentrations. It is around 44$^{\circ}$C at 4\,mm distance from the jet nozzle in the free
flowing effluent when 0.6$\%$ of O$_2$ are admixed. The gas temperature at the surface placed at
4\,mm distance from the nozzle is smaller then this value as verified by the treatment of skin,
which does not cause any pain or discomfort.

\begin{figure}
\centering
\includegraphics[width=8cm]{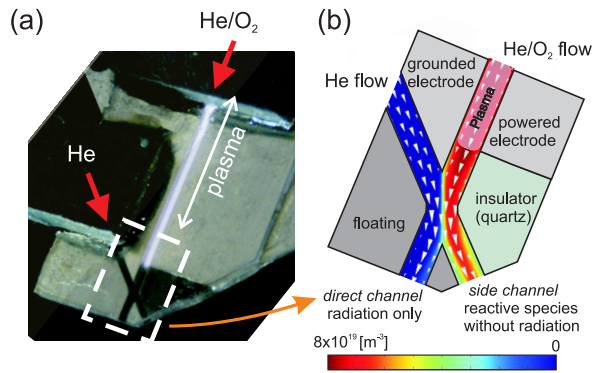}
\caption{The modification of the source geometry. Additional helium flow diverts the plasma
effluent in a side channel. VUV and UV photons propagate in line-of-sight with the plasma through
the direct channel. (a) Photograph of the X-Jet operated with He/O$_2$ plasma. (b) Results of 2D
fluid simulation of gas flow and O atom reaction kinetics; arrows correspond to gas velocity field,
the color map represents the simulated density of O atoms.}\label{Fig:XJet}
\end{figure}

The transport of atoms and molecules (heavy particles) in the plasma jets with substantial gas
flows of several liters/min under atmospheric pressure conditions is mainly controlled by
convection. The typical diffusion times from a $\sim$1\,mm diameter gas stream are comparable or
longer than the transport time due to convection, which is typically few hundred microseconds. On
the other hand, the plasma generated photons propagate in the direction of line-of-sight regardless
of the gas flow in the plasma effluent. This difference can be used to separate spatially the VUV
and UV photons generated in the plasma from heavy particles in the effluent. An additional flow of
helium gas crossing the plasma effluent with comparable or larger flow rate will divert plasma
effluent in another direction, while the photons will propagate undisturbed. However, care has to
be taken to prevent formation of turbulent flow and to avoid admixture of ambient atmosphere (air)
into the flow. This admixture could result in uncontrolled absorption of VUV and UV photons as well
as in the change of ROS densities due to reactions of radicals or metastables with O$_2$, N$_2$, or
H$_2$O molecules from air. To prevent this admixture we have extended the nozzle of the $\mu$-APPJ
from Fig.\ref{fig:APPJ} by two crossed channels as shown in a photograph in Fig.\ref{Fig:XJet}(a).
We call this jet modification an X-Jet and the two channels \emph{direct channel} and \emph{side
channel}. The plasma channel is extended as the \emph{direct channel} beyond the electrodes and is
crossed under a 45 degree angle by the \emph{side channel} with the same 1x1\,mm$^2$ cross section.
Additional He flow is applied to the \emph{side channel} to divert the plasma effluent from the
\emph{direct channel} into the \emph{side channel}. VUV and UV emission, on the other hand,
propagates further through the \emph{direct channel} (also filled with He) and can be used for
surface treatment. This emission contains all wavelengths emitted by the plasma including for
example the He excimer continuum emission at 58-100\,nm range \cite{Kurunczi2001}. This excimer
emission is blocked, when a MgF$_2$ filter (cutting wavelength 115\,nm) is used to separate the VUV
photons from heavy particles. The role of VUV radiation is often neglected in the discussion of the
results obtained by atmospheric pressure plasmas. Alternatively, the treatment by the emission
passed through a quartz window (cutting wavelength around 180 nm) is tested or only emission
measured in the 200-1000\,nm wavelength range was taken into account \cite{Perni2007}. The above
mentioned He$_2^*$ excimer continuum or strong atomic oxygen lines at 98 and 115\,nm are neglected
in these cases.

\section{Substrate preparation}

A plasma polymer film has been used in etching experiments as a model layer of the organic
material. We have used a hydrogenated amorphous carbon (a-C:H) film, which was deposited on a glass
substrate in an inductively coupled plasma reactor \cite{vonKeudell2010} from 2\,Pa of CH$_4$ gas
at power of 200\,W pulsed at 500\,Hz with a duty cycle of 10$\%$. The a-C:H film had 45 atomic
percent of hydrogen, density of 1\,g/cm$^3$ and the carbon was mainly sp$^3$ bonded. The film
thickness was 300\,nm with root mean square surface roughness of 3.8\,nm. The etched profile was
recorded by a profilometer (Dek Tak Stylus Profiler 6M) by measuring a line across the center of
the etched area. The same line was measured before etching (baseline measurement) to determine the
curvature of the surface due to bending of the glass substrate. This baseline is subtracted from
the measured etched profile.

Vegetative \emph{B. subtilis} and \emph{E. coli} cells have been treated by both types of the
microplasma jets. The following procedure has been used to prepare and evaluate them. \emph{B.
subtilis} 168 and \emph{E. coli} K12 liquid cultures were incubated for 18 hours over night at
37$^{\circ}$C in LB medium \cite{Sambrook1989}. The cultures were diluted to an optical density of
0.1 at 500\,nm (\emph {B. subtilis}) or 580\,nm (\emph {E. coli}) and were sprayed for 1 second
onto LB agar plates. A monolayer surface coverage with $\sim$4.5$\times$10$^{3}$cm$^{-2}$ cell
density is achieved in this case. The plates were grown for 2 hours at 37$^{\circ}$C before plasma
treatment. After plasma treatment, the sample plates were incubated over night for 18 hours at
37$^{\circ}$C to allow survivors to growth. Zone of inhibitions were observed where treatment was
lethal. The diameter of the zone of inhibition is used as a simple measure of the effect of plasma
on the bacteria. The shape of the zone of inhibition varies from circular to elliptic and the
diameter is reported as an average of its major and minor diameters. Additionally, we have checked
that treating of the agar plates with plasma before the application of cells had no effect on
bacterial growth. Furthermore, no pH change of the medium occurred during plasma treatment.

\section{Fluid model of plasma effluent}

A fluid model of the gas flow combined with the basic kinetic model of the reaction of O atoms with
O$_2$ molecules has been created and solved (using commercial COMSOL 3.5 software) to get better
insight into processes in the plasma effluent and to estimate the flux of atomic oxygen to the
surface. Two geometries have been simulated. First, the crossing of the channels in the X-Jet has
been modelled in a 2D geometry assuming for simplicity infinite thickness of electrodes. Second,
the 2D axially symmetric geometry has been used to simulate the oxygen chemistry in the effluent
and the flux of oxygen atoms to the surface at the jet surface distance of 4\,mm. Governing
equations and boundary conditions, which have been used in both models, will be discussed now.

The flow of He is described by incompressible momentum conservation and continuity equations:

\begin{eqnarray}
\quad \rho \frac{\partial\textbf{u}}{\partial t} - \nabla \cdot\eta(\nabla\textbf{u} + (\nabla\textbf{u})^T) + \rho \textbf{u} + \nabla p = 0 \\
\quad \nabla \cdot \textbf{u} = 0 \label{Model_F}
\end{eqnarray}

with $\rho$ being He gas density (0.164\,kg/m$^3$ at 101325 Pa and 300 K), $u$ gas flow velocity,
$\eta$ dynamic viscosity of He (2$\times$10$^{-5}$ Pa/s), and $p$ the pressure (101325 Pa). No
volume force is assumed to work on the gas. As boundary conditions no slip has been selected for
walls, the average velocity calculated from the gas flow and cross section of the channel has been
applied to the gas inlets ($\sim$25.6 m/s for 1.4\,slm He flow through 1\,mm$^2$) and a constant
pressure condition ($p$ = 101325 Pa) has been chosen for any outlet boundary. 300\,K gas
temperature is assumed in the model.

The transport of O and O$_3$ species has been simulated as a diffusion-convection transport with
diffusion coefficients of 1.29$\times$10$^{-4}$ m$^2$s$^{-1}$ and 0.713$\times$10$^{-4}$
m$^2$s$^{-1}$ for O and O$_3$, respectively (taken from ref. \cite{Waskoenig2010}) and with
velocities taken from the He flow simulation. The governing equation for species $i$ (i = O, O$_3$)
is:

\begin{equation}
\frac{\partial n_i}{\partial t} + \nabla \cdot(-D_i\nabla n_i) = R_i - \textbf{u} \nabla n_i
\label{Model_D}
\end{equation}

The $R_i$ is the production or loss of particles due to gas phase reactions. The volume in both
geometries is divided into two parts. A plasma region, where steady state conditions with constant
densities of all species are assumed, and effluent, where the reactive species recombine. For
simplicity, no reactions are assumed in the plasma region (the densities of O and O$_3$ are taken
from the literature and introduced into the model via boundary condition) and only three body
reaction of atomic oxygen, O$_2$, and He is considered in the effluent.

\begin{equation}
\mathrm{O + O_2 + He} \rightarrow \mathrm{O_3 + He} \label{R_O}
\end{equation}

This is the dominant gas phase reaction in the plasma effluent, where negligible densities of ions
or electrons can be assumed \cite{Waskoenig2010}. Reaction rate constant of this reaction is $k =
3.4\times10^{-46} (300/T_g)^{1.2} m^6s^{-1}$ \cite{Stafford2004}. For simplicity, constant values
of $2.48\times10^{25} m^{-3}$ (101325 Pa) and $1.49\times10^{23} m^{-3}$ (0.6\% of 101325 Pa) are
taken for the helium and molecular oxygen, respectively. The following boundary conditions have
been used: the densities of O and O$_3$ are set to $8\times10^{20} m^{-3}$ and $3\times10^{20}
m^{-3}$, respectively, at the gas inlet of the plasma channel. These values are extrapolated values
from the experimental measurements in the effluent \cite{Ellerweg2010}. The convection flow is
assumed for the outlet boundaries. The boundary conditions at any wall are different for O$_3$ and
O. Ozone is assumed to be unreactive at the wall with insulation/symmetry boundary condition at any
solid surface. Oxygen can on the contrary recombine or react at the surface, which has to be
considered in the model. The Neumann's boundary condition is therefore used at any solid surface
with normal flux of O atoms being lost at that boundary defined as \cite{Chantry87}:

\begin{equation}
F_{O,surface} = -n_O \cdot \frac{1}{2} \cdot \frac{\beta}{(2-\beta)} \cdot v
\label{Model_D}
\end{equation}

where the minus sign indicates the loss process, $\beta$ is the surface reaction probability of O
atoms at a given surface, and $v$ is the gas thermal velocity given by $(8k_BT/\pi m_i)^{0.5}$. The
surface reaction probability $\beta$ depends on the surface. It can be very low for example for
glass (10$^{-5}$ to 10$^{-3}$, \cite{Cartry99}) and could be close to unity for very reactive
surfaces. We take $\beta_{glass}$ = 10$^{-3}$ for glass surface of the channels and $\beta_{a-C:H}$
= 0.03 for the a-C:H surface. The variation of the former one in the above mentioned range does not
modify significantly the results of the simulation and the choice of the latter one will be
explained later in the text. The results of the simulation for the crossing of channels in the
X-Jet are shown in Fig.\ref{Fig:XJet}(b). The gas flow direction and velocity magnitude is
indicated by the white arrows (maximum gas velocity on the axis of a single channel is around
33.8\,m/s) and the color map represents the simulated density of atomic oxygen. It can be seen that
majority of oxygen atoms are diverted into the \emph{side channel} of the X-Jet. Only small amounts
diffuse into the \emph{direct channel}. This diffusion can be prevented by using higher He flow
through the \emph{side channel}. But we will demonstrate later that even with both He flows being
the same the flux of any heavy reactive particle from the plasma to the surface under the
\emph{direct channel} is negligible.

\begin{figure}
\centering
\includegraphics[width=8cm]{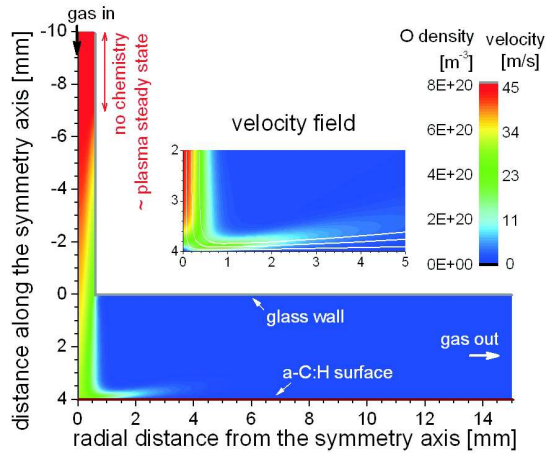}
\caption{The results of the 2D axially symmetric fluid model. The large color map represents the
simulated O density in the gas channel (region from -10 to 0 mm on the y-axis with first 3\,mm
being simulated without reaction (\ref{R_O})) and in the gap between the jet and substrate (0 to 4
mm on the y-axis). Y-axis is the symmetry axis of the model and boundary conditions (glass wall,
a-C:H surface, gas inlet and gas outlet) are indicated. The small inset shows the velocity field
and streamlines in the zoomed region near the substrate.}\label{Fig:model}
\end{figure}

The flux of atomic oxygen towards the surface can be modelled better with the 2D axially symmetric
geometry. This model does not allow to model both channels at the same time. However, it describes
more correctly the gas flow in a single channel (gas flow is surrounded by walls at each side) and
after the channel nozzle. The square 1$\times$1\,mm$^2$ channel is approximated by a cylinder with
the radius of 0.564 mm giving the same cross section area. Fig.\ref{Fig:model} shows the geometry
selected for the simulation. First, a 10\,mm long cylinder represents the last 3\,mm of plasma
channel (no gas phase reaction simulating the plasma steady state) followed by 7\,mm of gas flow in
a channel without a plasma (taking into account the O recombination reaction (\ref{R_O})). These
7\,mm represent the transport in one of the channels of the X-Jet. The effluent outside the channel
is simulated as a 4\,mm high cylinder with radius of 15\,mm. Both top and bottom boundaries are
assumed to be walls for simplicity. The top one is assumed to have properties of glass and the
bottom one is assumed to be an a-C:H film with different O atom surface reaction probability. The
outer wall of this cylinder serves as an outlet boundary in this case. The atomic oxygen and ozone
densities in the volume and the flux of atomic oxygen to the surface can be estimated with this
model.

The color map in Fig.\ref{Fig:model} shows the atomic oxygen density in the gas channel and in the
region near the surface. The inset in Fig.\ref{Fig:model} shows  the velocity field in the region
where the gas stream hits the a-C:H surface. Additionally, three streamlines, along which the gas
flow moves, are also shown in the inset. The prediction of the model can be validated by comparison
with the experimental data. The densities of O and O$_3$ have been measured as function of the
distance from the jet nozzle for the experimental conditions simulated here. The MBMS and TALIF
measurements show that O density decreases to 50$\%$ at the distance of 11\,mm from the nozzle
\cite{Ellerweg2010}. The simulation predicts a decrease to 65$\%$ after transport through the 7\,mm
tube and 4\,mm gap. The simulated decrease is smaller than the measured one. However, it has to be
taken into account that measurements were done with the free-standing effluent, without confining
glass tube around it. The flow velocity drops without the glass tube confining the gas flow.
Additionally, losses of O atoms due to the radial diffusion (which is hindered by the glass surface
of the tube) will also cause the faster decrease of O density in the effluent. We performed the
same simulation with "free-standing" effluent without confining glass tube and the O density
decreases to 55$\%$ of its original value, in a very good agreement with the measured data. It
corroborates that reaction (\ref{R_O}) is the dominant reaction in the plasma effluent. The
simulation also predicts that the majority of O atoms reacts with O$_2$ and is transformed into
O$_3$ molecules. This result is also in agreement with the experimental observations, where the
drop of O density is accompanied by a similar increase of O$_3$ density \cite{Ellerweg2010}.

The model can help us now to understand the processes governing the spacial distribution of O
densities in the effluent and to determine the O flux towards the surface. The O density is highest
on the jet axis, where the gas velocity is the largest and the transport time from the plasma is
the shortest. That is why the O density drops only to 65$\%$ of its original value in plasma. The
larger the distance from the axis, the slower is the gas flow and the faster is the drop of the O
density due to the three body gas phase reaction with O$_2$ and He. The gas flow is redirected at
the surface from the axial movement into the radial movement. The streamlines in
Fig.\ref{Fig:model} show that mainly the gas from the axis and near axis region, where the O
density is the highest, comes close to the substrate surface. That is why the surface reactions at
the glass wall inside the jet or some impurities entering the gas channel near the wall, as it is
the case in the \emph{direct channel} of the X-Jet with additional He gas flow, do not play any
significant role in the surface processes at the treated substrate. No significant change of the O
flux to the surface is observed even if the surface reaction probability of O at the glass wall is
set to 1. The O density in the gap between the jet and the substrate is high only at those places,
where also the gas velocity is large (cf the distribution of the O density and the velocity field
in Fig.\ref{Fig:model}). It is again the consequence of fast gas phase reaction of O with O$_2$.
The O density is large only in those regions, where the transport time from the plasma is shorter
or comparable with the reaction time of reaction (\ref{R_O}). The surface loss of O atoms at the
substrate only has a marginal effect on the bulk O density here because the diffusion is slow
compared to gas phase reactions. Only a thin boundary layer just near the substrate is depleted if
the surface reaction probability of O at a-C:H film surface is very large. The reaction probability
will then influence the profile of the O flux to the substrate, as it is shown later in this
article.

\section{Experimental results and discussion}

The big advantage of the X-Jet is that it allows us to separate the effect of plasma generated VUV
and UV photons from the effects induced by ROS (O$_3$, O, impurities,...). The effects of VUV and
UV photons only, ROS only, and photons and ROS together on a given sample (for example vegetative
bacterial cells on agar plates) can be studied in the following ways: 1) radiation and ROS
together: An X-Jet without additional He flow in the \emph{side channel} will result in the
transport of both ROS and (V)UV through the direct channel. 2) ROS only: The same He flow is used
in both channels. The additional flow through the \emph{side channel} will push the heavy particles
from the plasma effluent into the \emph{side channel} as demonstrated in Fig.\ref{Fig:XJet}(b). The
flow rates through both channels after the crossing will be the same due to the symmetry of this
geometry. The flux of ROS at the exit of the \emph{side channel} should be similar to the ROS flux
at the direct channel in case 1). Some differences will occur due to a missing photo-dissociation
and excitation of ROS and O$_2$ after the crossing of both channels (see also discussion of
\emph{E. coli} treatment later) and due to a slightly asymmetric velocity field across the
\emph{side channel}. The plasma generated VUV and UV photons cannot enter directly into \emph{side
channel} due to geometry constrains, which was corroborated by measurements presented later in this
article. 3) VUV and UV only: With additional He flow, only VUV and UV photons without ROS are at
the exit of the \emph{direct channel}. Higher He flow in the \emph{side channel} can be used to
make sure that no ROS from the plasma diffuse into the \emph{direct channel}.

We report in the following results of several tests, which were carried out to demonstrate the
performance of the X-Jet. Afterwards, results of treating bacteria with the unmodified $\mu$-APPJ
source in air followed by results of treating \emph{E. coli} with the X-Jet in controlled He
atmosphere will be presented and discussed.

\subsection{Effectiveness of separation of heavy particles and photons}

\emph{Measurements of emission intensity:} The light emission from the \emph{side channel} was
compared to the emission emitted through the direct channel. First, the emission spectrum in the
200-875\,nm wavelength range was measured (Ocean Optics USB4000 spectrometer, data not shown). The
atomic oxygen lines at 777.4 and 844.7\,nm dominated the spectrum, the third most intens line was
the He line at 706.6\,nm. Additionally, other He emission lines and emission band of OH around
309\,nm were observed in the \emph{direct channel}. Only the most intens 777.4 line was observed on
axis of the \emph{side channel} with intensity at 0.3$\%$ of the value observed in the \emph{direct
channel}. Slightly higher intensity (around 0.7$\%$) is observed slightly off axis of the
\emph{side channel} under the angle of $\sim$45 degree to the channel axis. This measurements show
that the light emission through the \emph{side channel} is reduced to less than 1$\%$ of the value
in the \emph{direct channel} and that reflections inside the channel structure are probably
responsible for the residual emission. However, there is still a possibility that the VUV photons
with wavelength below 200\,nm will be emitted by some long-living excited species, or that VUV
photons from the plasma will be reabsorbed and reemitted in the direction of the \emph{side
channel}. Therefore, the solar blind VUV and UV detector (PMT-142, effective in 115 - 450 nm
wavelength range with maximum relative efficiency at around 220\,nm\cite{Bibinov1997}) in the
evacuated housing with MgF$_2$ window has been used to measure the intensity of VUV and UV emission
from both channels of the X-Jet and also from the $\mu$-APPJ directly. A 1\,mm diameter diaphragm
was placed on the MgF$_2$ window and the jet was always at 4\,mm distance from the window to
maintain the same acceptance angle for each measurement. The same experimental conditions with
0.6$\%$ of O$_2$ were used again. Again, the emission from the \emph{side channel} was
significantly reduced, now to 0.7$\%$ on axis and to 1.5$\%$ off axis of the \emph{side channel}.
No equipment for measuring the complete emission spectrum including wavelengths below 115\,nm is
currently available in our lab so we cannot exclude that some emission at short wavelength range is
emitted through the \emph{side channel}. However, the fact that the emission above 115\,nm was
reduced below at least 1.5$\%$ and the observation, which will be shown and discussed later, that
VUV and UV radiation from the \emph{direct channel} has only a very small effect on bacteria,
strongly indicate that VUV and UV radiation through the \emph{side channel} is negligible.
Additionally, the signal intensity of $\mu$-APPJ and that of the \emph{direct channel} of the X-Jet
could be compared. The installation of the channel structure to the $\mu$-APPJ reduced the emission
intensity at 4\,mm distance from the jet to approximately 17 percent.

\begin{figure}
\centering
\includegraphics[width=8cm]{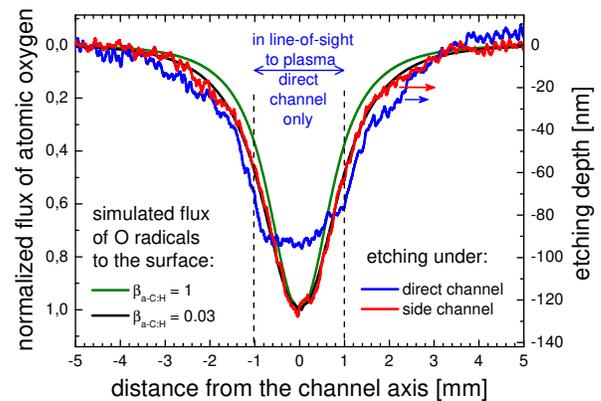}
\caption{Etched profiles generated in the model a-C:H film after 4\,min of the combined treatment
(\emph{direct channel}) and ROS-only treatment (\emph{side channel}) with He/O$_2$ plasma (0.6$\%$
O$_2$ in He). Applied RMS voltage was 230\,V. The profiles were measured across the middle of the
treated region. Additionally, a simulated normal flux of atomic oxygen to the surface is shown for
two surface reaction probabilities of O at the a-C:H surface, $\beta_{a-C:H}$ = 1 and
$\beta_{a-C:H}$ = 0.03. Dashed lines indicate, which part of the surface is exposed to the VUV and
UV photons in the case of the combined treatment.}\label{Fig:etchprofiles}
\end{figure}

\emph{Treatment of a model a-C:H film:} A model polymer film, the plasma deposited a-C:H layer, was
treated with different components of the plasma effluent separated in the X-Jet.
Fig.\ref{Fig:etchprofiles} shows the etching profiles induced by the treatment with the \emph{side
channel} and with additional He flow (further on called ROS-only treatment) and by the \emph{direct
channel} without additional He flow (further on called combined treatment). The gas channel used
for the treatment was positioned perpendicular to the treated surface with the channel
exit-substrate distance of 4\,mm. These experiments were performed in He atmosphere to suppress
possible admixture of surrounding air. The treatment time was 4\,min with 0.6$\%$ of O$_2$ in the
gas mixture and RMS voltage of 230\,V.

The etching profile after the ROS-only treatment was bell shaped with a maximum etch depth of
120\,nm and a full width at half maximum of around 2\,mm. It extended up to $\sim$4\,mm from its
symmetry axis (diameter $\sim$8\,mm). This is an expected etching profile caused by reactive
species like atomic oxygen, who's density in the gas flow decreases with time (and therefore also
with the distance from the jet axis) due to gas phase reactions and surface recombination. The flux
of atomic oxygen to the substrate was also simulated by the 2D axially symmetric simulation as
described above. The resulting normal flux of this reactive particles for two surface reaction
probabilities of O at the a-C:H surface, $\beta_{a-C:H}$ = 1 and $\beta_{a-C:H}$ = 0.03, are
compared to the etched profiles in Fig.\ref{Fig:etchprofiles}. It can be seen that the shape of the
etched profile in the ROS-only treatment case matches perfectly the simulated flux of atomic oxygen
to the surface when $\beta_{a-C:H}$ = 0.03 is used, which is a reasonable value between
$\beta_{glass}$ and 1. However, the profile simulated with $\beta_{a-C:H}$ = 1 is also very
similar, just with a little bit smaller width. The small effect of $\beta_{a-C:H}$ on the profile
is due to the fact that the fast gas phase reaction (\ref{R_O}) determines the O density near the
surface. Higher $\beta_{a-C:H}$ makes therefore only a small difference in the O depletion in the
boundary layer close to the a-C:H surface. Additionally, the predicted absolute flux of
8$\times$10$^{16}$ of O atoms/cm$^2$/s is higher than the measured etching rate of carbon atoms of
2.8$\times$10$^{15}$ of C atoms/cm$^2$/s (calculated based on a-C:H film density of 1 kg/m$^3$ and
45 atomic percent of hydrogen in the film). It can therefore explain the observed etch rate very
well even assuming that two O atoms are needed per C to form CO$_2$ and that some O atoms recombine
with each other or react with hydrogen at the surface. We conclude therefore that most probably the
atomic oxygen is responsible for the etching of the model a-C:H film and that the area reached by
atomic oxygen has a diameter of no more than 10\,mm.

The etching profile looked different after 4 minutes of the combined treatment (\emph{direct
channel}, X-Jet without additional He flow). The etched profile was similar only at a distance
greater than 1\,mm from the jet axis. However, the etching just underneath the jet, where the a-C:H
film was in direct line-of-sight to plasma, was significantly slower than in the previous case.
This difference in the etching profiles could be explained by hardening of the model a-C:H film by
VUV and UV photons. The quanta of VUV and UV radiation have enough energy to break C-H and C-C
bonds in the a-C:H film resulting in enhanced cross-linking and densification of the material,
which makes it more difficult for atomic oxygen to etch it. The data presented in
Fig.\ref{Fig:etchprofiles} nicely demonstrates how important it is to consider VUV and UV radiation
generated by the plasma in these experiments. The change of the surface profile of a-C:H film after
the treatment with VUV and UV photons only (\emph{direct channel}, X-Jet with additional He flow)
has been checked as well and no etching has been observed after 4 or 20 minutes. This observation
corroborates that O atoms do not reach the treated surface in this case even if the same fluxes of
He are used in both channels, the case in which some ROS and O$_2$ diffuse into the \emph{direct
channel} at the channel crossing. However, these species will not diffuse up to the axis of the
\emph{direct channel} and will, therefore, not be transported to the vicinity of the treated
substrate (see the discussion of the 2D axially symmetric model and streamlines in
Fig.\ref{Fig:model}).

\subsection{Treatment of bacteria with $\mu$-APPJ}

\begin{figure}
\centering
\includegraphics[width=8cm]{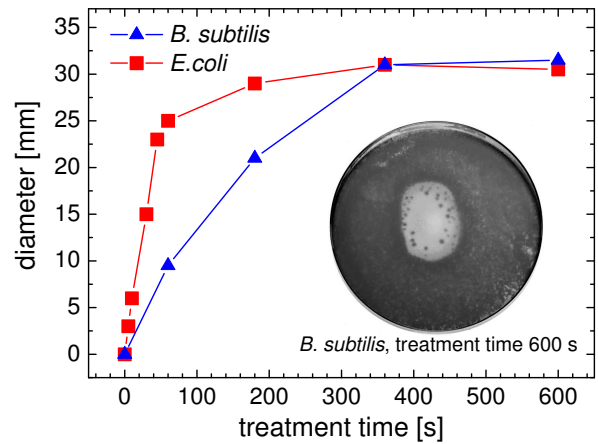}
\caption{Diameter of zone of inhibition for \emph{B. subtilis} and \emph{E. coli} as function of
the treatment time as induced by the $\mu$-APPJ. Plasma parameters: 1.4 slm He with 8.4 sccm O$_2$
(concentration 0.6\%), U$_{RMS}$=230\,V, jet-substrate distance 4\,mm, ambient atmosphere.
Photograph shows the size of a zone of inhibition for \emph{B. subtilis} after 600\,s of the
treatment.}\label{Fig:Hemmhofen}
\end{figure}

The bacteria on agar plates were treated with the effluent of the $\mu$-APPJ source in ambient
atmosphere. The jet was at 4\,mm distance with He/O$_2$ gas mixture (0.6$\%$ O$_2$ in He) and RMS
voltage of 230\,V. The densities of the most dominant ROS species, atomic oxygen, and ozone, are
around 6$\times$10$^{14}$ cm$^{-3}$ at 4\,mm distance under these conditions \cite{Ellerweg2010}.
The plasma emission is localized between the electrodes and there is no direct contact of this
active plasma region with the substrate. Fig.\ref{Fig:Hemmhofen} shows the diameter of the zone of
inhibition as a function of treatment time. Additionally, the photograph shows the zone of
inhibition for \emph{B. subtilis} after 600\,s of the treatment. The diameter is measured up to the
region with the homogeneous density of bacteria.

For both bacteria, the diameter of the zone of inhibition reached a final value of 30\,mm
(Fig.\ref{Fig:Hemmhofen}). It indicates that the maximal distance on the surface from the jet axis,
which the reactive species responsible for inactivation can reach (in the ambient air atmosphere)
is 15\,mm. We have already shown that the effective distance the atomic oxygen can reach was less
than 5\,mm. The flux of VUV and UV photons is the most intense just underneath the jet but will be
absorbed by air at any radial distance from the jet axis larger than a few millimeters. Therefore,
some long-living reactive species have to be responsible for bacterial inactivation at 15\,mm
distance. We propose that ozone is this reactive species. It is known for its bactericidal effects.
Just 5\,min treatment with 0.2\,ppm of ozone in water is lethal for \emph{E. coli}, \emph{B.
cereus}, or \emph{B. megaterium} \cite{Broadwater1973}. Moreover, experiments in air with ozone
have shown that it is also effective in killing bacteria laying on the agar plates. Ozone
concentrations below 1\,ppm and treatment times less than 100\,min have been reported to be
effective in killing \emph{Staphylococcus albus}, \emph{Streptococcus salivarius}, and \emph{B.
prodigiosus} \cite{Elford1942}. Ozone is produced effectively in the jet with concentrations at the
surface higher than 20\,ppm, well above the lethal limits reported in the literature. However, this
does not explain why the zone of inhibition is limited to 30\,mm diameter only. Moreover, the
experiments performed in helium atmosphere, which are discussed in the next paragraph, do not
exhibit this behavior. For example, the whole Petri dish with 80\,mm diameter was affected after 3
minutes of combined treatment.

\begin{figure}
\centering
\includegraphics[width=5.5cm]{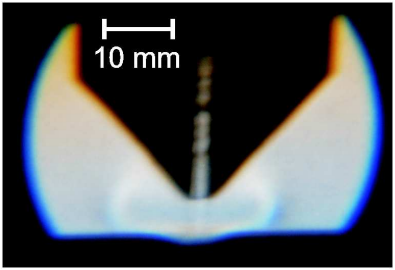}
\caption{Schlieren image of the $\mu$-APPJ with 1.4\,slm of He (no O$_2$ added, plasma off)
pointing downwards perpendicularly to the plane at 4\,mm distance.}\label{Fig:Schlieren}
\end{figure}

These observations can be explained as follows. Ozone is responsible for the inactivation at large
distances from the jet. It is lethal in the concentrations generated in the jet and it is stable
enough to survive the transport. The difference between the treatment in air and in He atmosphere
is the buoyancy force acting on the He/O$_2$/O$_3$ gas mixture in the surrounding air. The O$_2$
and O$_3$ concentrations are so small that the average density of the mixture is very close to He
density and the diffusion is slow preventing fast depletion of O$_2$ and O$_3$ densities from He
and fast mixing with air. The buoyancy force lifts up the He/O$_2$/O$_3$ flow from the probe
surface at some distance from the jet axis. Ozone is lifted up with the He flow and can therefore
not reach the surface at a distance larger than 15\,mm from the jet axis. The experimental evidence
for this hypothesis is shown in Fig.\ref{Fig:Schlieren}, which presents a schlieren image of the
$\mu$-APPJ with 1.4\,slm He gas flow pointing downwards to the plane at 4\,mm distance. The
brighter and darker regions mark the presence of density gradients revealing in this case the
boundary between He and air. It can be seen, that this boundary is around 12\,mm from the jet axis.
It corroborates that the He flow is indeed lifted from the surface. Moreover, this boundary moves a
few millimeters over time indicating that turbulent or unstable flow takes place in this region.
These observations can very well explain the maximum diameter of the zone of inhibition in the
treatment with air as ambient atmosphere.

The buoyancy force is not present when He is used as the surrounding atmosphere and the flow of
plasma effluent can spread undisturbed along the substrate surface. Ozone can therefore reach and
inactivate the whole area of the Petri dish in this case.

Fig.\ref{Fig:Hemmhofen} also shows that the increase of diameter with the treatment time is 4 to 5
times faster for \emph{E. coli} than for \emph{B. subtilis}. The reason probably is that \emph{B.
subtilis}, as a Gram-positive bacterium, has a thicker cell wall and is protected better against
the plasma treatment. \emph{B. subtilis} is well known for its resistance against environmental
stresses.

\subsection{Treatment of bacteria with X-Jet}

\begin{figure}
\centering
\includegraphics[width=8cm]{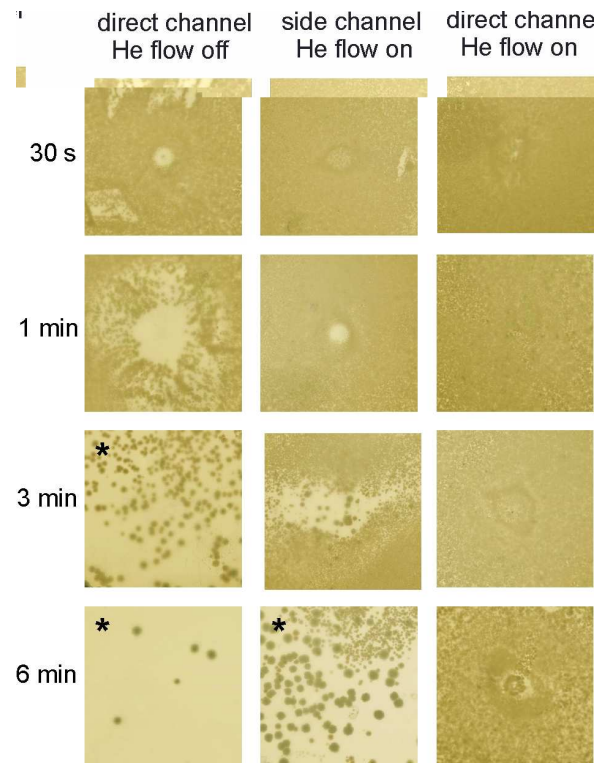}
\caption{40 by 40 mm details of photographs of Petri dishes with zones of inhibition after 30\,s
and 1, 3, and 6\,min of combined, ROS-only, and VUV and UV photons only treatments of \emph{E.
coli} monolayers. Details marked by $\ast$: the entire plate (80\,mm diameter) was
affected.}\label{Fig:EColi}
\end{figure}

\emph{E. coli} cells were exposed to the same treatment as a-C:H films. Fig.\ref{Fig:EColi} shows
the resulting zone of inhibitions after 0.5, 1, 3, and 6 minutes of combined, ROS-only and
radiation only treatments of \emph{E. coli}. Again, the treatments were performed in a closed
reactor filled with He. Because the O$_2$ gas injected into the jet is depleted only up to several
percent, the reactor is filled with a He/O$_2$ gas mixture. We used an additional He gas flow into
the reactor to further reduce the O$_2$ concentration in the background atmosphere to below
0.2\,$\%$. The second constituent of the background atmosphere with larger concentration was O$_3$,
which is produced in the jet.

The VUV and UV only experiment (\emph{direct channel}, X-Jet with additional He flow) showed that
these photons induce only a very small effect. No inactivation was observed up to 3\,min of the
treatment and two small zones of inhibition appeared after 6\,min of treatment. One was located
directly on the jet axis (diameter $<$1.5\,mm) and was probably caused directly by VUV and UV
photons. The other one was donut-shaped with a diameter of about 6\,mm. This latter one was
probably caused by admixture of O$_2$ or O$_3$ from the background gas into the effluent, their
photo dissociation and reaction at the surface.

The combined treatment and ROS-only treatment showed typical dose-effect relationships. Elongated
treatment times resulted in larger zones of inhibition and lower numbers of colony forming units.
Contrary to the experiments in ambient air, where the maximum size of the zone of inhibition was
limited to 30\,mm, the whole area of the Petri dish is affected after 3 and 6\,min of combined
treatment and after 6\,min of ROS-only treatment. The 40 by 40\,mm areas (marked by asterisk in
Fig.\ref{Fig:EColi}) are representative for the whole area of the Petri dishes (diameter 80\,mm) in
these three cases. As already discussed above, ozone is the most probable candidate for the
inactivation at large distance from the jet axis. VUV and UV photons generated in the plasma reach
only a very small area ($\sim$2\,mm diameter) directly under the plasma jet and reactive atomic
oxygen is depleted quickly from the gas phase as it was shown by the etching experiments and by the
2D axially symmetric simulation.

Surprisingly, the effect of the combined treatment was approximately twice as fast as the effect of
the ROS-only treatment (cf Fig.\ref{Fig:EColi}). We expected to see a synergistic effect between
ROS and photons directly under the jet, where cells are exposed to both simultaneously. The
diameter of this area is $\sim$2\,mm only. We also expected that the effect outside of this area
will be similar for both the combined and ROS-only treatments, in a similar way to that observed
for the etching of a-C:H films. This, however, was not the case. The zone of inhibition of the
combined treatment had a diameter of 5\,mm after 30\,s and the density of cells was decreased up to
20\,mm from the jet axis after 1\,min of treatment. The ROS-only treatment showed no zone of
inhibition after 30\,s and the visibly affected area only had a 5\,mm diameter after a 1\,min
treatment. Additionally, the combined treatment at 3\,min looks similar to the ROS-only treatment
at 6\,min. These observations indicate that the plasma effluent is changed by the presence of VUV
and UV photons, some photochemistry takes place, and more reactive or excited species reach the
substrate. Further investigations are planned to better understand this effect.

\section{Conclusions}

The modification of the microscale atmospheric pressure plasma jet ($\mu$-APPJ) operated in
He/O$_2$, so called X-Jet, has been used to test the possibility of separating VUV and UV photons
emitted by plasma from reactive particles transported by the gas flow in the plasma effluent. This
separation provides the opportunity to study separately the effects of these components on living
organisms, biological macromolecules or model films. We have shown that emission intensity of
photons in the 115-875\,nm wavelength region is reduced to less than 1.5$\%$ in the gas channel
transporting the reactive oxygen species and we have demonstrated the performance of the X-Jet in
etching experiments using a model a-C:H film as a substrate. Similar etching profiles were observed
for both combined treatment and for ROS-only treatment. However, in the case of the combined
treatment, the VUV and UV photons cause hardening of the model film on the area in line-of-sight to
the plasma, which results into slower etch rates directly under the jet nozzle. The fluid model
simulation of the gas flows and O reaction kinetics indicate that O atoms are responsible for the
etching of the polymer film and that they can reach an area with maximum diameter of 10\,mm.

The plasma effluent has been used for the treatment of bacteria.  First, we have verified that the
$\mu$-APPJ can inactivate \emph{B. subtilis} and \emph{E. coli}. Treatment of several minutes leads
to zones of inhibition with a diameter of 30\,mm with faster inactivation of \emph{E. coli}.
Second, \emph{E. coli} monolayers on agar plates were treated by the X-Jet in a controlled helium
atmosphere. Under the conditions tested, the VUV and UV photons alone had only a small effect on
the bacteria. In the ROS-only and combined treatments, the cells were most probably inactivated by
ozone at larger distances from the jet and by combined effect of ozone, atomic oxygen, and some
other possible impurities in the region close to the jet axis. We could show that the size of zone
of inactivation is limited by the buoyancy force when working in ambient air. Furthermore, the
results indicate that the VUV and UV photons can produce effectively reactive species in the gas
phase, which leads to faster inactivation of \emph{E. coli}.

We believe that the X-Jet is a powerful tool for the fundamental study of inactivation mechanisms
by atmospheric pressure plasmas. Furthermore, this jet can also be used for any surface processing,
which requires high fluxes of ROS without VUV and UV radiation, such as treatment of polymers or
living tissues.

\section{Acknowledgement}
The authors thank Volker Schultz-von der Gathen for fruitful discussions about the operation of the
$\mu$-APPJ source and Henrik B\"{o}ttner and Nick Knake for the help with Schlieren imaging of the
jet. This work has been performed with the support of the research group FOR1123 approved by the
German Research Foundation (DFG). This work has also been supported by the Research Department
Plasmas with Complex Interactions of the Ruhr-Universit\"{a}t Bochum.


\section{References}

\begin{thebibliography}{32}
\expandafter\ifx\csname natexlab\endcsname\relax\def\natexlab#1{#1}\fi \expandafter\ifx\csname
bibnamefont\endcsname\relax
  \def\bibnamefont#1{#1}\fi
\expandafter\ifx\csname bibfnamefont\endcsname\relax
  \def\bibfnamefont#1{#1}\fi
\expandafter\ifx\csname citenamefont\endcsname\relax
  \def\citenamefont#1{#1}\fi
\expandafter\ifx\csname url\endcsname\relax
  \def\url#1{\texttt{#1}}\fi
\expandafter\ifx\csname urlprefix\endcsname\relax\def\urlprefix{URL }\fi
\providecommand{\bibinfo}[2]{#2} \providecommand{\eprint}[2][]{\url{#2}}

\bibitem[1]{Kong2009}
\bibinfo{author}{\bibfnamefont{M.~G.} \bibnamefont{Kong}},
  \bibinfo{author}{\bibfnamefont{G.}~\bibnamefont{Kroesen}},
  \bibinfo{author}{\bibfnamefont{G.}~\bibnamefont{Morfill}},
  \bibinfo{author}{\bibfnamefont{T.}~\bibnamefont{Nosenko}},
  \bibinfo{author}{\bibfnamefont{T.}~\bibnamefont{Shimizu}},
  \bibinfo{author}{\bibfnamefont{J.}~\bibnamefont{van Dijk}}, \bibnamefont{and}
  \bibinfo{author}{\bibfnamefont{J.~L.} \bibnamefont{Zimmermann}},
  \bibinfo{journal}{New J. Phys.} \textbf{\bibinfo{volume}{11}},
  \bibinfo{pages}{115012} (\bibinfo{year}{2009}).

\bibitem[2]{Stoffels2008}
\bibinfo{author}{\bibfnamefont{E.}~\bibnamefont{Stoffels}},
  \bibinfo{author}{\bibfnamefont{Y.}~\bibnamefont{Sakiyama}}, \bibnamefont{and}
  \bibinfo{author}{\bibfnamefont{D.~B.} \bibnamefont{Graves}},
  \bibinfo{journal}{IEEE Trans. Plasma Sci.} \textbf{\bibinfo{volume}{36}},
  \bibinfo{pages}{1441} (\bibinfo{year}{2008}).

\bibitem[3]{Daeschlein2010}
\bibinfo{author}{\bibfnamefont{G.}~\bibnamefont{Daeschlein}},
  \bibinfo{author}{\bibfnamefont{T.}~\bibnamefont{von Woedtke}},
  \bibinfo{author}{\bibfnamefont{E.}~\bibnamefont{Kindel}},
  \bibinfo{author}{\bibfnamefont{R.}~\bibnamefont{Brandenburg}},
  \bibinfo{author}{\bibfnamefont{K.-D.} \bibnamefont{Weltmann}},
  \bibnamefont{and}
  \bibinfo{author}{\bibfnamefont{M.}~\bibnamefont{J\"{u}nger}},
  \bibinfo{journal}{Plasma Process. Polym.} \textbf{\bibinfo{volume}{7}},
  \bibinfo{pages}{224} (\bibinfo{year}{2010}{\natexlab{a}}).

\bibitem[4]{Daeschlein2010a}
\bibinfo{author}{\bibfnamefont{G.}~\bibnamefont{Daeschlein}},
  \bibinfo{author}{\bibfnamefont{S.}~\bibnamefont{Scholz}},
  \bibinfo{author}{\bibfnamefont{T.}~\bibnamefont{von Woedtke}},
  \bibinfo{author}{\bibfnamefont{M.}~\bibnamefont{Niggemeier}},
  \bibinfo{author}{\bibfnamefont{E.}~\bibnamefont{Kindel}},
  \bibinfo{author}{\bibfnamefont{R.}~\bibnamefont{Foest}},
  \bibinfo{author}{\bibfnamefont{R.}~\bibnamefont{Brandenburg}},
  \bibinfo{author}{\bibfnamefont{K.-D.} \bibnamefont{Weltmann}},
  \bibnamefont{and}
  \bibinfo{author}{\bibfnamefont{M.}~\bibnamefont{J\"{u}nger}},
  \bibinfo{journal}{IEEE Trans. Plasma Sci.} p. \bibinfo{pages}{DOI:
  10.1109/TPS.2010.2063441} (\bibinfo{year}{2010}{\natexlab{b}}).

\bibitem[5]{Deng2007}
\bibinfo{author}{\bibfnamefont{X.~T.} \bibnamefont{Deng}},
  \bibinfo{author}{\bibfnamefont{J.~J.} \bibnamefont{Shi}}, \bibnamefont{and}
  \bibinfo{author}{\bibfnamefont{M.~G.} \bibnamefont{Kong}},
  \bibinfo{journal}{J. Appl. Phys.} \textbf{\bibinfo{volume}{101}},
  \bibinfo{pages}{074701} (\bibinfo{year}{2007}).

\bibitem[6]{Fridman2007}
\bibinfo{author}{\bibfnamefont{G.}~\bibnamefont{Fridman}},
  \bibinfo{author}{\bibfnamefont{A.}~\bibnamefont{Shereshevsky}},
  \bibinfo{author}{\bibfnamefont{M.}~\bibnamefont{Jost}},
  \bibinfo{author}{\bibfnamefont{A.}~\bibnamefont{Brooks}},
  \bibinfo{author}{\bibfnamefont{A.}~\bibnamefont{Fridman}},
  \bibinfo{author}{\bibfnamefont{A.}~\bibnamefont{Gutsol}},
  \bibinfo{author}{\bibfnamefont{V.}~\bibnamefont{Vasilets}}, \bibnamefont{and}
  \bibinfo{author}{\bibfnamefont{G.}~\bibnamefont{Friedman}},
  \bibinfo{journal}{Plasma Chem. Plasma Process.}
  \textbf{\bibinfo{volume}{27}}, \bibinfo{pages}{163} (\bibinfo{year}{2007}).

\bibitem[7]{Vandamme2010}
\bibinfo{author}{\bibfnamefont{M.}~\bibnamefont{Vandamme}},
  \bibinfo{author}{\bibfnamefont{E.}~\bibnamefont{Robert}},
  \bibinfo{author}{\bibfnamefont{S.}~\bibnamefont{Pesnel}},
  \bibinfo{author}{\bibfnamefont{E.}~\bibnamefont{Barbosa}},
  \bibinfo{author}{\bibfnamefont{S.}~\bibnamefont{Dozias}},
  \bibinfo{author}{\bibfnamefont{J.}~\bibnamefont{Sobilo}},
  \bibinfo{author}{\bibfnamefont{S.}~\bibnamefont{Lerondel}}, \bibnamefont{and}
  \bibinfo{author}{\bibfnamefont{A.}~\bibnamefont{Le Pape}},
  \bibinfo{author}{\bibfnamefont{J.~M.}~\bibnamefont{Pouvesle}},
  \bibinfo{journal}{Plasma Process. Polym.} \textbf{\bibinfo{volume}{7}},
  \bibinfo{pages}{264} (\bibinfo{year}{2010}).

\bibitem[8]{Dobrynin2009}
\bibinfo{author}{\bibfnamefont{D.}~\bibnamefont{Dobrynin}},
  \bibinfo{author}{\bibfnamefont{G.}~\bibnamefont{Fridman}},
  \bibinfo{author}{\bibfnamefont{G.}~\bibnamefont{Friedman}}, \bibnamefont{and}
  \bibinfo{author}{\bibfnamefont{A.}~\bibnamefont{Fridman}},
  \bibinfo{journal}{New J. Phys.} \textbf{\bibinfo{volume}{11}},
  \bibinfo{pages}{115020} (\bibinfo{year}{2009}).

\bibitem[9]{Morfill2009}
\bibinfo{author}{\bibfnamefont{G.}~\bibnamefont{Morfill}},
  \bibinfo{author}{\bibfnamefont{T.}~\bibnamefont{Shimizu}},
  \bibinfo{author}{\bibfnamefont{B.}~\bibnamefont{Steffes}}, \bibnamefont{and}
  \bibinfo{author}{\bibfnamefont{H.-U.} \bibnamefont{Schmidt}},
  \bibinfo{journal}{New J. Phys.} \textbf{\bibinfo{volume}{11}},
  \bibinfo{pages}{115019} (\bibinfo{year}{2009}).

\bibitem[10]{Cao2009}
\bibinfo{author}{\bibfnamefont{Z.}~\bibnamefont{Cao}},
  \bibinfo{author}{\bibfnamefont{Q.}~\bibnamefont{Nie}},
  \bibinfo{author}{\bibfnamefont{D.~L.} \bibnamefont{Bayliss}},
  \bibinfo{author}{\bibfnamefont{J.~L.} \bibnamefont{Walsh}},
  \bibinfo{author}{\bibfnamefont{C.~S.} \bibnamefont{Ren}},
  \bibinfo{author}{\bibfnamefont{D.~Z.} \bibnamefont{Wang}}, \bibnamefont{and}
  \bibinfo{author}{\bibfnamefont{M.~G.} \bibnamefont{Kong}},
  \bibinfo{journal}{Plasma Sources Sci. Technol.}
  \textbf{\bibinfo{volume}{19}}, \bibinfo{pages}{025003}
  (\bibinfo{year}{2009}).

\bibitem[11]{Laroussi2005}
\bibinfo{author}{\bibfnamefont{M.}~\bibnamefont{Laroussi}} \bibnamefont{and}
  \bibinfo{author}{\bibfnamefont{X.}~\bibnamefont{Lu}}, \bibinfo{journal}{Appl.
  Phys. Lett.} \textbf{\bibinfo{volume}{87}}, \bibinfo{pages}{113902}
  (\bibinfo{year}{2005}).

\bibitem[12]{Robert2009}
\bibinfo{author}{\bibfnamefont{E.}~\bibnamefont{Robert}},
  \bibinfo{author}{\bibfnamefont{E.}~\bibnamefont{Barbosa}},
  \bibinfo{author}{\bibfnamefont{S.}~\bibnamefont{Dozias}},
  \bibinfo{author}{\bibfnamefont{M.}~\bibnamefont{Vandamme}},
  \bibinfo{author}{\bibfnamefont{C.}~\bibnamefont{Cachoncinlle}},
  \bibinfo{author}{\bibfnamefont{R.}~\bibnamefont{Viladrosa}},
  \bibnamefont{and} \bibinfo{author}{\bibfnamefont{J.~M.}
  \bibnamefont{Pouvesle}}, \bibinfo{journal}{Plasma Process. Polym}
  \textbf{\bibinfo{volume}{6}}, \bibinfo{pages}{795} (\bibinfo{year}{2009}).

\bibitem[13]{Schulz-vonderGathen2007}
\bibinfo{author}{\bibfnamefont{V.}~\bibnamefont{{Schulz-von der Gathen}}},
  \bibinfo{author}{\bibfnamefont{V.}~\bibnamefont{Buck}},
  \bibinfo{author}{\bibfnamefont{T.}~\bibnamefont{Gans}},
  \bibinfo{author}{\bibfnamefont{N.}~\bibnamefont{Knake}},
  \bibinfo{author}{\bibfnamefont{K.}~\bibnamefont{Niemi}},
  \bibinfo{author}{\bibfnamefont{S.}~\bibnamefont{Reuter}},
  \bibinfo{author}{\bibfnamefont{L.}~\bibnamefont{Schaper}}, \bibnamefont{and}
  \bibinfo{author}{\bibfnamefont{J.}~\bibnamefont{Winter}},
  \bibinfo{journal}{Contr. Plasma Phys.} \textbf{\bibinfo{volume}{47}},
  \bibinfo{pages}{510} (\bibinfo{year}{2007}).

\bibitem[14]{Ehlbeck2011}
\bibinfo{author}{\bibfnamefont{J.}~\bibnamefont{Ehlbeck}},
  \bibinfo{author}{\bibfnamefont{U.}~\bibnamefont{Schnabel}},
  \bibinfo{author}{\bibfnamefont{M.}~\bibnamefont{Polak}},
  \bibinfo{author}{\bibfnamefont{J.}~\bibnamefont{Winter}},
  \bibinfo{author}{\bibfnamefont{T.}~\bibnamefont{von Woedtke}},
  \bibinfo{author}{\bibfnamefont{R.}~\bibnamefont{Brandenburg}},
  \bibinfo{author}{\bibfnamefont{T.}~\bibnamefont{von~dem Hagen}},
  \bibnamefont{and} \bibinfo{author}{\bibfnamefont{K.-D.}
  \bibnamefont{Weltmann}}, \bibinfo{journal}{J. Phys. D: Appl. Phys.}
  \textbf{\bibinfo{volume}{44}}, \bibinfo{pages}{013002}
  (\bibinfo{year}{2011}).

\bibitem[15]{Goree2006}
\bibinfo{author}{\bibfnamefont{J.}~\bibnamefont{Goree}},
  \bibinfo{author}{\bibfnamefont{B.}~\bibnamefont{Liu}}, \bibnamefont{and}
  \bibinfo{author}{\bibfnamefont{D.}~\bibnamefont{Drake}}, \bibinfo{journal}{J.
  Phys. D: Appl. Phys.} \textbf{\bibinfo{volume}{39}}, \bibinfo{pages}{3479}
  (\bibinfo{year}{2006}).

\bibitem[16]{Haehnel2010}
\bibinfo{author}{\bibfnamefont{M.}~\bibnamefont{H\"{a}hnel}},
  \bibinfo{author}{\bibfnamefont{T.}~\bibnamefont{von Woedtke}},
  \bibnamefont{and} \bibinfo{author}{\bibfnamefont{K.-D.}
  \bibnamefont{Weltmann}}, \bibinfo{journal}{Plasma Process. Polym.}
  \textbf{\bibinfo{volume}{7}}, \bibinfo{pages}{244} (\bibinfo{year}{2010}).

\bibitem[17]{Jeong98}
\bibinfo{author}{\bibfnamefont{J.}~\bibnamefont{Jeong}},
  \bibinfo{author}{\bibfnamefont{S.}~\bibnamefont{Babayan}},
  \bibinfo{author}{\bibfnamefont{A.}~\bibnamefont{Sch\"utze}},
  \bibinfo{author}{\bibfnamefont{V.}~\bibnamefont{Tu}},
  \bibinfo{author}{\bibfnamefont{M.}~\bibnamefont{Morajev}},
  \bibinfo{author}{\bibfnamefont{G.}~\bibnamefont{Selwyn}}, \bibnamefont{and}
  \bibinfo{author}{\bibfnamefont{R.}~\bibnamefont{Hicks}},
  \bibinfo{journal}{Plasma Sources Sci. Technol.} \textbf{\bibinfo{volume}{7}},
  \bibinfo{pages}{282} (\bibinfo{year}{1998}).

\bibitem[18]{Laimer2006}
\bibinfo{author}{\bibfnamefont{J.}~\bibnamefont{Laimer}} \bibnamefont{and}
  \bibinfo{author}{\bibfnamefont{H.}~\bibnamefont{St\"ori}},
  \bibinfo{journal}{Plasma Process. Polym.} \textbf{\bibinfo{volume}{3}},
  \bibinfo{pages}{573} (\bibinfo{year}{2006}).

\bibitem[19]{Waskoenig2010}
\bibinfo{author}{\bibfnamefont{J.}~\bibnamefont{Waskoenig}},
  \bibinfo{author}{\bibfnamefont{K.}~\bibnamefont{Niemi}},
  \bibinfo{author}{\bibfnamefont{N.}~\bibnamefont{Knake}},
  \bibinfo{author}{\bibfnamefont{L.~M.} \bibnamefont{Graham}},
  \bibinfo{author}{\bibfnamefont{S.}~\bibnamefont{Reuter}},
  \bibinfo{author}{\bibfnamefont{V.}~\bibnamefont{{Schulz-von der Gathen}}},
  \bibnamefont{and} \bibinfo{author}{\bibfnamefont{T.}~\bibnamefont{Gans}},
  \bibinfo{journal}{Plasma Sources Sci. Technol.}
  \textbf{\bibinfo{volume}{19}}, \bibinfo{pages}{045018}
  (\bibinfo{year}{2010}).

\bibitem[20]{Liu2010}
\bibinfo{author}{\bibfnamefont{D.-X.} \bibnamefont{Liu}},
  \bibinfo{author}{\bibfnamefont{M.-Z.} \bibnamefont{Rong}},
  \bibinfo{author}{\bibfnamefont{X.-H.} \bibnamefont{Wang}},
  \bibinfo{author}{\bibfnamefont{F.}~\bibnamefont{Iza}},
  \bibinfo{author}{\bibfnamefont{M.~G.} \bibnamefont{Kong}}, \bibnamefont{and}
  \bibinfo{author}{\bibfnamefont{P.}~\bibnamefont{Bruggeman}},
  \bibinfo{journal}{Plasma Process. Polym.} \textbf{\bibinfo{volume}{7}},
  \bibinfo{pages}{846} (\bibinfo{year}{2010}).

\bibitem[21]{Knake08}
\bibinfo{author}{\bibfnamefont{N.}~\bibnamefont{Knake}},
  \bibinfo{author}{\bibfnamefont{S.}~\bibnamefont{Reuter}},
  \bibinfo{author}{\bibfnamefont{K.}~\bibnamefont{Niemi}},
  \bibinfo{author}{\bibfnamefont{V.}~\bibnamefont{{Schulz-von der Gathen}}},
  \bibnamefont{and} \bibinfo{author}{\bibfnamefont{J.}~\bibnamefont{Winter}},
  \bibinfo{journal}{J. Phys. D: Appl. Phys.} \textbf{\bibinfo{volume}{41}},
  \bibinfo{pages}{194006} (\bibinfo{year}{2008}).

\bibitem[22]{Ellerweg2010}
\bibinfo{author}{\bibfnamefont{D.}~\bibnamefont{Ellerweg}},
  \bibinfo{author}{\bibfnamefont{J.}~\bibnamefont{Benedikt}},
  \bibinfo{author}{\bibfnamefont{A.} \bibnamefont{von Keudell}},
  \bibinfo{author}{\bibfnamefont{N.}~\bibnamefont{Knake}},
  \bibnamefont{and} \bibinfo{author}{\bibfnamefont{V.}~\bibnamefont{{Schulz-von
  der Gathen}}}, \bibinfo{journal}{New J. Phys.} \textbf{\bibinfo{volume}{12}},
  \bibinfo{pages}{013021} (\bibinfo{year}{2010}).

\bibitem[23]{Kurunczi2001}
\bibinfo{author}{\bibfnamefont{P.}~\bibnamefont{Kurunczi}},
  \bibinfo{author}{\bibfnamefont{J.}~\bibnamefont{Lopez}},
  \bibinfo{author}{\bibfnamefont{H.}~\bibnamefont{Shah}}, \bibnamefont{and}
  \bibinfo{author}{\bibfnamefont{K.}~\bibnamefont{Becker}},
  \bibinfo{journal}{Int. J. Mass Spec.} \textbf{\bibinfo{volume}{205}},
  \bibinfo{pages}{277} (\bibinfo{year}{2001}).

\bibitem[24]{Perni2007}
\bibinfo{author}{\bibfnamefont{S.}~\bibnamefont{Perni}},
  \bibinfo{author}{\bibfnamefont{G.}~\bibnamefont{Shama}},
  \bibinfo{author}{\bibfnamefont{J.~L.} \bibnamefont{Hobman}},
  \bibinfo{author}{\bibfnamefont{P.~A.} \bibnamefont{Lund}},
  \bibinfo{author}{\bibfnamefont{C.~J.} \bibnamefont{Kershaw}},
  \bibinfo{author}{\bibfnamefont{G.~A.} \bibnamefont{Hidalgo-Arroyo}},
  \bibinfo{author}{\bibfnamefont{C.~W.} \bibnamefont{Penn}},
  \bibinfo{author}{\bibfnamefont{X.~T.} \bibnamefont{Deng}},
  \bibinfo{author}{\bibfnamefont{J.~L.} \bibnamefont{Walsh}}, \bibnamefont{and}
  \bibinfo{author}{\bibfnamefont{M.~G.} \bibnamefont{Kong}},
  \bibinfo{journal}{Applied Physics Letters} \textbf{\bibinfo{volume}{90}},
  \bibinfo{pages}{073902} (\bibinfo{year}{2007}).

\bibitem[25]{vonKeudell2010}
\bibinfo{author}{\bibfnamefont{A.}~\bibnamefont{von Keudell}},
  \bibinfo{author}{\bibfnamefont{P.}~\bibnamefont{Awakowicz}},
  \bibinfo{author}{\bibfnamefont{J.}~\bibnamefont{Benedikt}},
  \bibinfo{author}{\bibfnamefont{V.}~\bibnamefont{Raballand}},
  \bibinfo{author}{\bibfnamefont{A.}~\bibnamefont{Yanguas-Gil}},
  \bibinfo{author}{\bibfnamefont{J.}~\bibnamefont{Opretzka}},
  \bibinfo{author}{\bibfnamefont{C.}~\bibnamefont{Fl\"otgen}},
  \bibinfo{author}{\bibfnamefont{R.}~\bibnamefont{Reuter}},
  \bibinfo{author}{\bibfnamefont{L.}~\bibnamefont{Byelykh}},
  \bibinfo{author}{\bibfnamefont{H.}~\bibnamefont{Halfmann}},
  \bibinfo{author}{\bibfnamefont{K.}~\bibnamefont{Stapelmann}},
  \bibinfo{author}{\bibfnamefont{B.}~\bibnamefont{Denis}},
  \bibinfo{author}{\bibfnamefont{J.}~\bibnamefont{Wunderlich}},
  \bibinfo{author}{\bibfnamefont{P.}~\bibnamefont{Muranyi}},
  \bibinfo{author}{\bibfnamefont{F.}~\bibnamefont{Rossi}},
  \bibinfo{author}{\bibfnamefont{O.}~\bibnamefont{Kyli\'an}},
  \bibinfo{author}{\bibfnamefont{N.}~\bibnamefont{Hasiwa}},
  \bibinfo{author}{\bibfnamefont{A.}~\bibnamefont{Ruiz}},
  \bibinfo{author}{\bibfnamefont{H.}~\bibnamefont{Rauscher}},
  \bibinfo{author}{\bibfnamefont{L.}~\bibnamefont{Sirghi}},
  \bibinfo{author}{\bibfnamefont{E.}~\bibnamefont{Comoy}},
  \bibinfo{author}{\bibfnamefont{C.}~\bibnamefont{Dehen}},
  \bibinfo{author}{\bibfnamefont{L.}~\bibnamefont{Challier}}, \bibnamefont{and}
  \bibinfo{author}{\bibfnamefont{J.~P.}~\bibnamefont{Deslys}},
  \bibinfo{journal}{Plasma Process. Polym.}
  \textbf{\bibinfo{volume}{7}}, \bibinfo{pages}{327} (\bibinfo{year}{2010}).

\bibitem[26]{Sambrook1989}
\bibinfo{author}{\bibfnamefont{J.}~\bibnamefont{Sambrook}},
  \bibinfo{author}{\bibfnamefont{E.}~\bibnamefont{Fritsch}}, \bibnamefont{and}
  \bibinfo{author}{\bibfnamefont{T.}~\bibnamefont{Maniatis}},
  \emph{\bibinfo{title}{Molecular Cloning: A Laboratory Manual, Volume 1 to 3,
  2nd edition}} (\bibinfo{publisher}{Cold Spring Harbor Laboratory Press},
  \bibinfo{year}{1989}).

\bibitem[27]{Stafford2004}
\bibinfo{author}{\bibfnamefont{D.~S.} \bibnamefont{Stafford}} \bibnamefont{and}
  \bibinfo{author}{\bibfnamefont{M.~J.} \bibnamefont{Kushner}},
  \bibinfo{journal}{J. Appl. Phys.} \textbf{\bibinfo{volume}{96}},
  \bibinfo{pages}{2451} (\bibinfo{year}{2004}).

\bibitem[28]{Chantry87}
\bibinfo{author}{\bibfnamefont{P.}~\bibnamefont{Chantry}}, \bibinfo{journal}{J.
  Appl. Phys.} \textbf{\bibinfo{volume}{62}}, \bibinfo{pages}{1141}
  (\bibinfo{year}{1987}).

\bibitem[29]{Cartry99}
\bibinfo{author}{\bibfnamefont{G.}~\bibnamefont{Cartry}},
  \bibinfo{author}{\bibfnamefont{L.}~\bibnamefont{Magne}}, \bibnamefont{and}
  \bibinfo{author}{\bibfnamefont{G.}~\bibnamefont{Cernogara}},
  \bibinfo{journal}{J. Phys. D} \textbf{\bibinfo{volume}{32}},
  \bibinfo{pages}{L53} (\bibinfo{year}{1999}).

\bibitem[30]{Bibinov1997}
\bibinfo{author}{\bibfnamefont{N.~K.} \bibnamefont{Bibinov}},
  \bibinfo{author}{\bibfnamefont{D.~O.} \bibnamefont{Bolshukhin}},
  \bibinfo{author}{\bibfnamefont{D.~B.} \bibnamefont{Kokh}},
  \bibinfo{author}{\bibfnamefont{A.~M.} \bibnamefont{Pravilov}},
  \bibinfo{author}{\bibfnamefont{I.~P.} \bibnamefont{Vinogradov}},
  \bibnamefont{and}
  \bibinfo{author}{\bibfnamefont{K.}~\bibnamefont{Wiesemann}},
  \bibinfo{journal}{Measurement Science and Technology}
  \textbf{\bibinfo{volume}{8}}, \bibinfo{pages}{773} (\bibinfo{year}{1997}).

\bibitem[31]{Broadwater1973}
\bibinfo{author}{\bibfnamefont{W.~T.} \bibnamefont{Broadwater}},
  \bibinfo{author}{\bibfnamefont{R.~C.} \bibnamefont{Hoehn}}, \bibnamefont{and}
  \bibinfo{author}{\bibfnamefont{P.~H.} \bibnamefont{King}},
  \bibinfo{journal}{Appl. Microbiol.} \textbf{\bibinfo{volume}{26}},
  \bibinfo{pages}{391} (\bibinfo{year}{1973}).

\bibitem[32]{Elford1942}
\bibinfo{author}{\bibfnamefont{W.~J.} \bibnamefont{Elford}} \bibnamefont{and}
  \bibinfo{author}{\bibfnamefont{J.}~\bibnamefont{{van den Ende}}},
  \bibinfo{journal}{J. Hygiene} \textbf{\bibinfo{volume}{42}},
  \bibinfo{pages}{240} (\bibinfo{year}{1942}).

\end{thebibliography}
\end{document}